\documentclass{PoS}
\usepackage{amsmath}
\usepackage{amssymb}
\usepackage{graphicx}
\usepackage{placeins}

\newcommand{\V}{{\textsl{v}}}

\renewcommand\refname{\vspace{-6.5ex}}

\renewenvironment{thebibliography}[1]{
  \begin{oldthebibliography}{#1}
    \setlength{\itemsep}{0.8ex}
    \setlength{\parskip}{0ex}
}
{
  \end{oldthebibliography}
}

\title{Radiative leptonic decays on the lattice}

\ShortTitle{Radiative leptonic decays on the lattice}

\author{Christopher Kane$^a$,
        Christoph Lehner$^{b,c}$,
        \speaker{Stefan Meinel}$^{\:a,d}$,
        Amarjit Soni$^c$\\

    $^{a}$Department of Physics, University of Arizona, Tucson, AZ 85721, USA\\
    $^{b}$Department of Physics, University of Regensburg, 93040 Regensburg, Germany\\
    $^{c}$Physics Department, Brookhaven National Laboratory, Upton, NY 11973, USA\\
    $^{d}$RIKEN BNL Research Center, Brookhaven National Laboratory, Upton, NY 11973, USA\\
\\ \\
E-mail: \email{smeinel@email.arizona.edu}
}

\abstract{Adding a hard photon to the final state of a leptonic pseudoscalar-meson decay lifts the helicity suppression and can provide sensitivity to a larger set of operators in the weak effective Hamiltonian. Furthermore, radiative leptonic $B$ decays at high photon energy are well suited to constrain the first inverse moment of the $B$-meson light-cone distribution amplitude, an important parameter in the theory of nonleptonic $B$ decays. We demonstrate that the calculation of radiative leptonic decays is possible using Euclidean lattice QCD, and present preliminary numerical results for $D_s^+ \to \ell^+ \nu\gamma\:\:$ and $\:\:K^- \to \ell^-\bar{\nu}\gamma$.
}

\FullConference{37th International Symposium on Lattice Field Theory - Lattice2019\\
		16-22 June 2019\\
		Wuhan, China}

\begin{document}

\begin{figure}[t]
 \includegraphics[height=0.105\textheight]{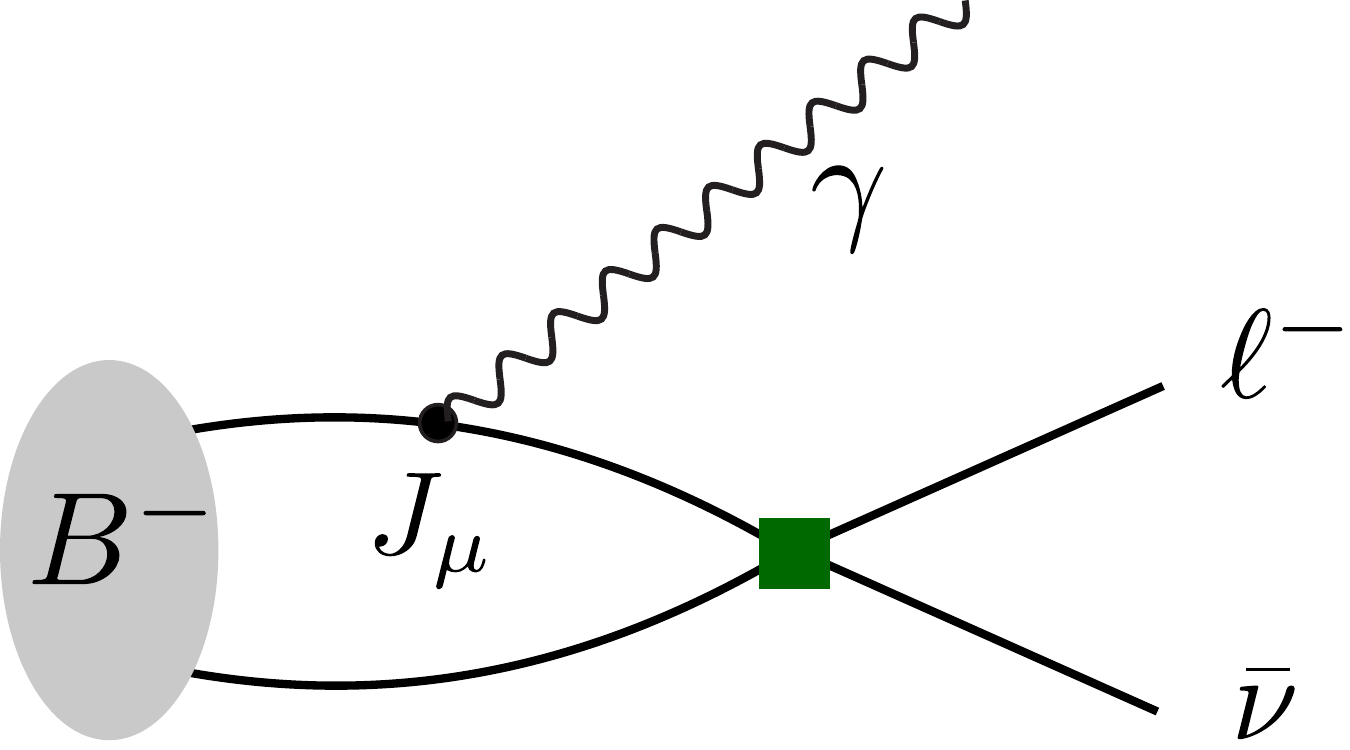} \hfill \includegraphics[height=0.105\textheight]{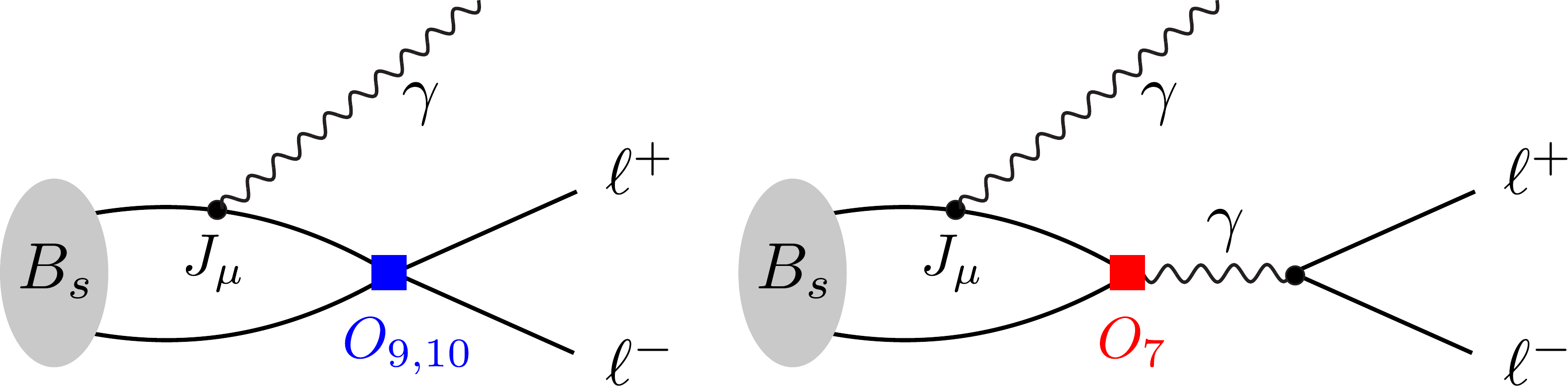}
 \caption{\label{fig:RLDs}Left: A diagram contributing to $B^- \to \ell^- \bar{\nu}\, \gamma$, where the green square corresponds to $W$-boson exchange in the Standard Model and comes with a factor of $V_{ub}$. Right: two diagrams contributing to
 $B_s \to \ell^+\ell^-\gamma$, via the operators $O_{7,9,10}$ (defined, for example, in Ref.~\cite{Aebischer:2019mlg}).}
\end{figure}

\section{Introduction}

Radiative leptonic decays of pseudoscalar mesons probe both the weak interaction and the hadronic structure in useful ways. Adding a sufficiently energetic photon to the final state can actually increase the branching fraction \cite{Atwood:1994za}, as it removes the helicity suppression. Perhaps the most interesting example is $B^- \to \ell^- \bar{\nu}\, \gamma$, shown in Fig.~\ref{fig:RLDs} (left). For large $E_\gamma^{(0)}$, this process is the cleanest probe of the first inverse moment of the $B$-meson light-cone distribution amplitude, $1/\lambda_B = \int_0^\infty\frac{\Phi_{B^+}(\omega)}{\omega}\:d\omega$, an important input in QCD-factorization predictions for nonleptonic $B$ decays that is presently poorly determined \cite{Korchemsky:1999qb,DescotesGenon:2002mw,Lunghi:2002ju,Braun:2012kp,Beneke:2018wjp,Wang:2018wfj,Beneke:1999br}. A recent search for this decay by Belle gave an upper limit $\mathcal{B}( B^- \to \ell^- \bar{\nu} \gamma,\:E_\gamma^{(0)}>1\:{\rm GeV} ) < 3.0\times 10^{-6}$, close to the Standard-Model expectation \cite{Gelb:2018end}. Lattice QCD results for the $B^- \to \ell^- \bar{\nu} \gamma$ form factors could be used to constrain $\lambda_B$. Also very interesting are the flavor-changing neutral-current decays  $B^0 \to \ell^+\ell^-\gamma$ and $B_s \to \ell^+\ell^-\gamma$ (shown in Fig.~\ref{fig:RLDs}, right). While the purely leptonic decays are sensitive to $C_{10,S,P}-C^\prime_{10,S,P}$ only, the radiative leptonic decays probe all Wilson coefficients in the weak effective Hamiltonian, including $C_9$, in which global fits of experimental results for other $b\to s\ell^+\ell^-$ decays indicate a deviation from the Standard Model that violates lepton flavor universality (LFU) \cite{Aebischer:2019mlg}. Since the radiative leptonic decays are not helicity-suppressed, they are well-suited for testing LFU with light leptons \cite{Guadagnoli:2017quo,Dettori:2016zff}. For the charmed-meson radiative
leptonic decays $D^+ \to e^+ \nu \gamma$ and $D_s^+ \to e^+ \nu \gamma$, the BESIII collaboration has reported upper limits on the branching fractions with $E_\gamma^{(0)} > 10\:{\rm MeV}$ of $3.0\times 10^{-5}$ and $1.3\times 10^{-4}$, respectively \cite{Ablikim:2017twd, Ablikim:2019bil}. Finally, in contrast to the heavy-meson decays, there are already precise measurements of the differential branching fractions of $K^- \to e^- \bar{\nu} \gamma$, $K^- \to \mu^- \bar{\nu} \gamma$, $\pi^- \to e^- \bar{\nu} \gamma$, and $\pi^- \to \mu^- \bar{\nu} \gamma$, as reviewed in Ref.~\cite{Tanabashi:2018oca}. These decay modes can therefore be used to test the lattice QCD methods.

In the following, we show how radiative leptonic decays can be calculated on a Euclidean lattice, and we present early numerical results. One of us previously reported on this project at the Lattice 2018 conference \cite{Soni2018}. At Lattice 2019, radiative leptonic decays were also discussed by G.~Martinelli \cite{Martinelli2019}.

\section{Hadronic tensor and form factors}

To define the form factors for charged-current radiative leptonic decays of pseudoscalar mesons, we use the notation for $B^- \to \ell^- \bar{\nu}\, \gamma$. The quark electromagnetic and weak currents are given by $J_\mu = \sum_q e_q\, \bar{q}\gamma_\mu q$ and $J^{\rm weak}_\nu = \bar{u}\gamma_\nu(1-\gamma_5) b$. The decay amplitude depends on the hadronic tensor, which is defined as
\begin{eqnarray}
 T_{\mu\nu} &=& -i \int d^4x \:\:e^{ip_\gamma\cdot x} \langle 0 | \mathsf{T} \left( J_\mu(x) \:J^{\rm weak}_\nu(0) \right) |B^-(\mathbf{p}_B) \rangle
\end{eqnarray}
in Minkowski space. Throughout this work, we asssume that the photon is real, i.e., $p_\gamma^2=0$. The hadronic tensor can be decomposed as \cite{Beneke:2018wjp}
\begin{eqnarray}
T_{\mu\nu} &=& \epsilon_{\mu\nu\tau\rho} p_\gamma^\tau \V^{\,\rho} {F_V} + i [-g_{\mu\nu} (p_\gamma \cdot \V) + \V_\mu (p_\gamma)_\nu ] {F_A} - i \frac{\V_\mu \V_\nu}{p_\gamma \cdot \V}  m_B {f_B} + (p_\gamma)_\mu\text{-terms}, \label{eq:FFdecomp}
\end{eqnarray}
where $p_B=m_B \V$ and the $(p_\gamma)_\mu\text{-terms}$ will disappear when contracting with the photon polarization vector. The form factors $F_V$ and $F_A$ are functions of the photon energy in the $B$-meson rest frame, $E_\gamma^{(0)} = p_\gamma\cdot\V =(m_B^2-q^2)/(2 m_B)$. Also appearing in Eq.~(\ref{eq:FFdecomp}) is the $B$-meson decay constant $f_B$.

To prepare for the discussion in the next section, it is useful to write down the spectral representation of $T_{\mu\nu}$ in Minkowski space for the two different time orderings of the currents. By inserting complete sets of energy/momentum eigenstates and performing the time integrals, we find
\begin{eqnarray}
\nonumber T^<_{\mu\nu} &=& -i  \int_{-\infty(1-i\epsilon)}^0\hspace{-2ex}dt\hspace{2ex}e^{iE_\gamma t} \int d^3x \:\:e^{-i\mathbf{p}_\gamma\cdot\mathbf{x}} \langle 0 |  J^{\rm weak}_\nu(0) \: J_\mu(t,\mathbf{x})  |B^-(\mathbf{p}_B) \rangle \\
 &=&  - \sum_n \frac{1}{2 E_{n,(\mathbf{p}_B-\mathbf{p}_\gamma)}} \frac{\langle 0 |  J^{\rm weak}_\nu(0) | n(\mathbf{p}_B-\mathbf{p}_\gamma) \rangle \langle n(\mathbf{p}_B-\mathbf{p}_\gamma) | J_\mu(0) | B(\mathbf{p}_B) \rangle}{E_\gamma + E_{n,(\mathbf{p}_B-\mathbf{p}_\gamma)}- E_B - i\epsilon} , \label{eq:Tless} \\
\nonumber T^>_{\mu\nu} &=& -i  \int_{0}^{\infty(1-i\epsilon)}\hspace{-2ex}dt\hspace{2ex}e^{iE_\gamma t} \int d^3x \:\:e^{-i\mathbf{p}_\gamma\cdot\mathbf{x}} \langle 0 |   J_\mu(t,\mathbf{x}) \: J^{\rm weak}_\nu(0)  |B^-(\mathbf{p}_B) \rangle \\ 
 &=&\sum_{m} \frac{1}{2 E_{m,\mathbf{p}_\gamma}} \frac{ \langle 0 |  J_\mu(0)| m(\mathbf{p}_\gamma) \rangle \langle m(\mathbf{p}_\gamma) | J^{\rm weak}_\nu(0)    | B(\mathbf{p}_B) \rangle}{E_\gamma-E_{m,\mathbf{p}_\gamma}-i\epsilon}  \label{eq:Tgreater}
\end{eqnarray}
(in infinite volume, the sums over $n$ and $m$ include integrals over the continuous spectrum of multi-particle states).

\section{Extracting the hadronic tensor from a Euclidean three-point function}

In this section, we show that $T_{\mu\nu}$ can be extracted from the Euclidean three-point function
\begin{eqnarray}
C_{\mu\nu} (t, t_B) &=& \int d^3x \int d^3y\:\: e^{-i\mathbf{p}_\gamma\cdot\mathbf{x}} e^{i\mathbf{p}_B\cdot\mathbf{y}} \left\langle J_\mu(t,\mathbf{x})\:\: J^{\rm weak}_\nu(0,\mathbf{0}) \:\: \phi_B^\dag(t_B,\mathbf{y})    \right\rangle,
\end{eqnarray}
where $\phi_B \sim \bar{u}\gamma_5 b$ is an interpolating field for the $B$ meson, and $t$, $t_B$ now denote the Euclidean time. We define the integrals
\begin{equation}
 I^<_{\mu\nu}(t_B, T) = \int_{-\!T}^0 dt\:\:e^{E_\gamma t}\:  C_{\mu\nu} (t, t_B), \hspace{5ex} I^>_{\mu\nu}(t_B, T) = \int_{0}^T dt\:\:e^{E_\gamma t}\:  C_{\mu\nu} (t, t_B),
\end{equation}
with a finite integration range $T$. Here we take $t_B$ to be large and negative (with $t_B < -T$), such that ground-state saturation is achieved for the $B$ meson. Inserting again complete sets of energy/momentum eigenstates, we find, for the first time ordering,
\begin{eqnarray}
\nonumber I^<_{\mu\nu}(t_B, T) &=& \langle B(\mathbf{p}_B)|\phi_B^\dag(0)|0\rangle \frac{1}{2 E_B} e^{E_B t_B}   \\
\nonumber && \times \sum_n \frac{1}{2 E_{n,(\mathbf{p}_B-\mathbf{p}_\gamma)}} \frac{ \langle 0 | J^{\rm weak}_\nu(0) | n(\mathbf{p}_B-\mathbf{p}_\gamma)\rangle \langle n(\mathbf{p}_B-\mathbf{p}_\gamma) | J_\mu(0) |B(\mathbf{p}_B) \rangle}{E_\gamma+E_{n,(\mathbf{p}_B-\mathbf{p}_\gamma)}-E_B} \\
&&\times  \left(1-e^{-(E_\gamma+E_{n,(\mathbf{p}_B-\mathbf{p}_\gamma)}-E_B) T}\right). \label{eq:Iless}
\end{eqnarray}
The sum over states in Eq.~(\ref{eq:Iless}) differs from the sum in Eq.~(\ref{eq:Tless}) by the factor in the last line. However, the exponential $e^{-(E_\gamma+E_{n,(\mathbf{p}_B-\mathbf{p}_\gamma)}-E_B) T}$ will vanish for large $T$ if $E_\gamma+E_{n,(\mathbf{p}_B-\mathbf{p}_\gamma)} > E_B$. Because the states $| n(\mathbf{p}_B-\mathbf{p}_\gamma)\rangle$ have the same quark-flavor quantum numbers as the $B$ meson, we have $E_{n,(\mathbf{p}_B-\mathbf{p}_\gamma)} \geq E_{B,(\mathbf{p}_B-\mathbf{p}_\gamma)}=\sqrt{m_B^2 + (\mathbf{p}_B-\mathbf{p}_\gamma)^2}$. Thus, we need $\displaystyle\sqrt{\mathbf{p}_\gamma^2} + \sqrt{m_B^2 + (\mathbf{p}_B-\mathbf{p}_\gamma)^2} > \sqrt{m_B^2 + \mathbf{p}_B^2}$. This is in fact always true if $\mathbf{p}_\gamma\neq 0$.

For the other time ordering, we find
\begin{eqnarray}
\nonumber  I^>_{\mu\nu}(t_B, T) &=& -\langle B(\mathbf{p}_B)|\phi_B^\dag(0)|0\rangle \frac{1}{2 E_B} e^{E_B t_B}   \\
 && \times \sum_m \frac{1}{2 E_{m,\mathbf{p}_\gamma}} \frac{\langle 0 | J_\mu(0) | m(\mathbf{p}_\gamma)\rangle \langle m(\mathbf{p}_\gamma) | J^{\rm weak}_\nu(0) |B(\mathbf{p}_B) \rangle}{E_\gamma-E_{m,\mathbf{p}_\gamma}} \left(1 - e^{(E_\gamma-E_{m,\mathbf{p}_\gamma}) T}\right).
\end{eqnarray}
The unwanted exponential $e^{(E_\gamma-E_{m,\mathbf{p}_\gamma}) T}$ in the last line goes to zero for large $T$ if $E_{m,\mathbf{p}_\gamma} > E_\gamma$.
Because the states $| m(\mathbf{p}_\gamma)\rangle$ are hadronic and have nonzero masses, their energies are larger than the energy of a photon with the same spatial momentum,
showing that this condition is also always satisfied. In summary, for $\mathbf{p}_\gamma \neq 0$,
\begin{eqnarray}
T_{\mu\nu} &=& -\lim_{T\to\infty}\:\: \lim_{t_B\to-\infty} \frac{2 E_B\,e^{-E_B t_B}}{\langle B(\mathbf{p}_B)|\phi_B^\dag(0)|0\rangle  }  I_{\mu\nu}(t_B, T),
\end{eqnarray}
where $I_{\mu\nu}$ is the integral from $-T$ to $T$. The energy $E_B$ and the overlap factor $\langle B(\mathbf{p}_B)|\phi_B^\dag(0)|0\rangle$ can be obtained from the two-point function $\int d^3 x\:\: e^{-i\mathbf{p}_B\cdot\mathbf{x}} \langle\phi_B(t,\mathbf{x})\: \phi_B^\dag(0)\rangle$.

Note that similar nonlocal matrix elements appear in processes with two photons, whose lattice calculation has been discussed, for example, in Refs.~\cite{Ji:2001wha,Dudek:2006ut,Feng:2012ck}.

\section{Preliminary numerical results}

\begin{figure}[b]

 \includegraphics[height=0.25\textheight]{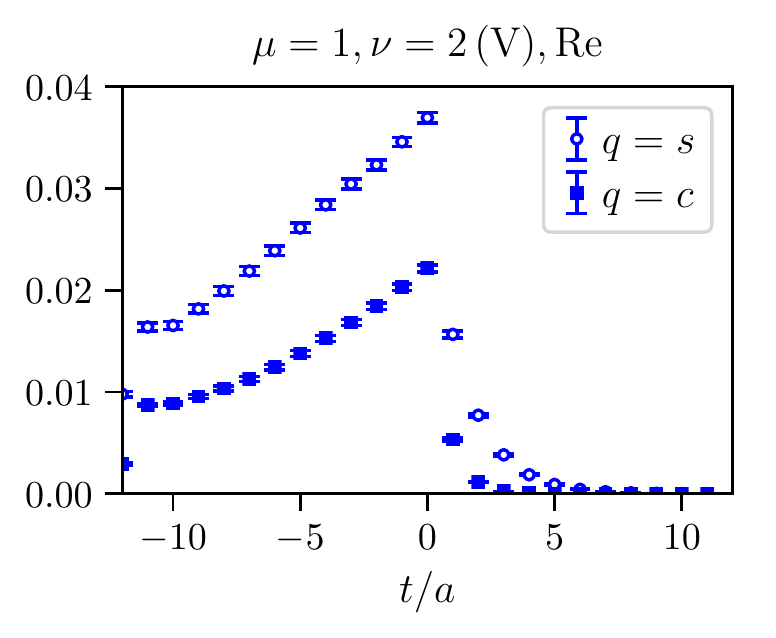} \hfill \includegraphics[height=0.25\textheight]{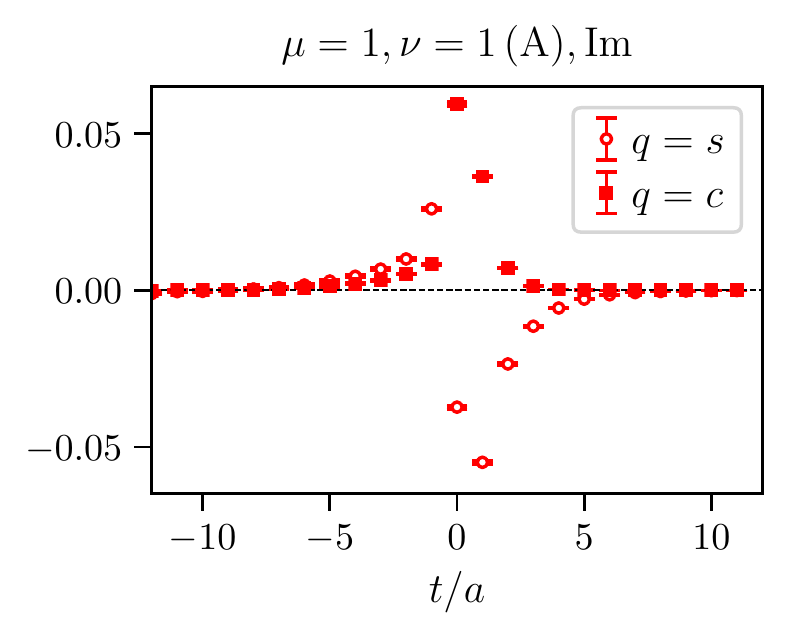}
 
 \vspace{-2ex}
 
 \caption{\label{fig:Cmunu}The unintegrated, scaled three-point functions $-\frac{2\,E_{D_s}\,e^{-E_{D_s} t_{D_s}}}{ \langle D_s(\mathbf{p}_{D_s})|\phi_{D_s}^\dag(0)|0\rangle  } C_{\mu\nu} (t, t_{D_s}) $ as a function of the electromagnetic-current insertion time $t$, for $t_{D_s}/a=-12$ and $\mathbf{p}_\gamma=(0,0,1)\frac{2\pi}{L}$. The left plot shows a combination of indices sensitive to $F_V$, while the right plot shows a combination sensitive to $F_A$. The contributions from the $s$ and $c$ quark in the electromagnetic current are shown separately, without charge factors.}
\end{figure}

\begin{figure}[t]

 \vspace{-3ex}

 \includegraphics[width=0.474\linewidth]{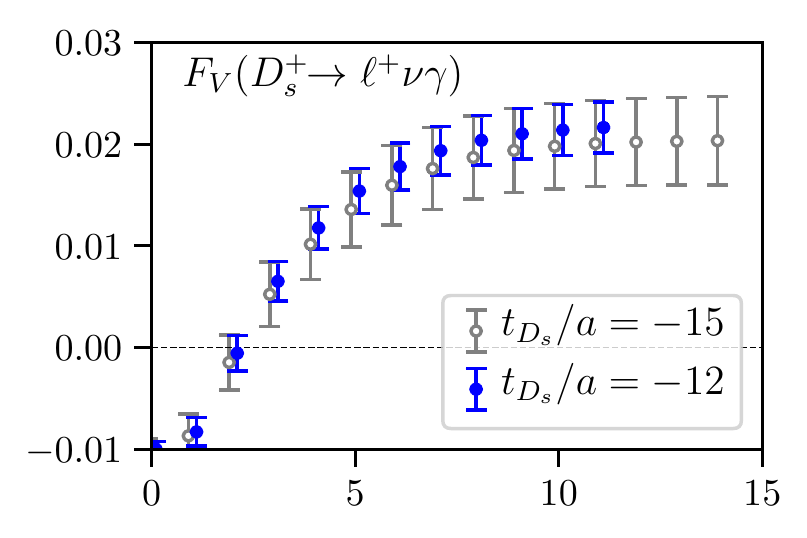} \hfill  \includegraphics[width=0.44\linewidth]{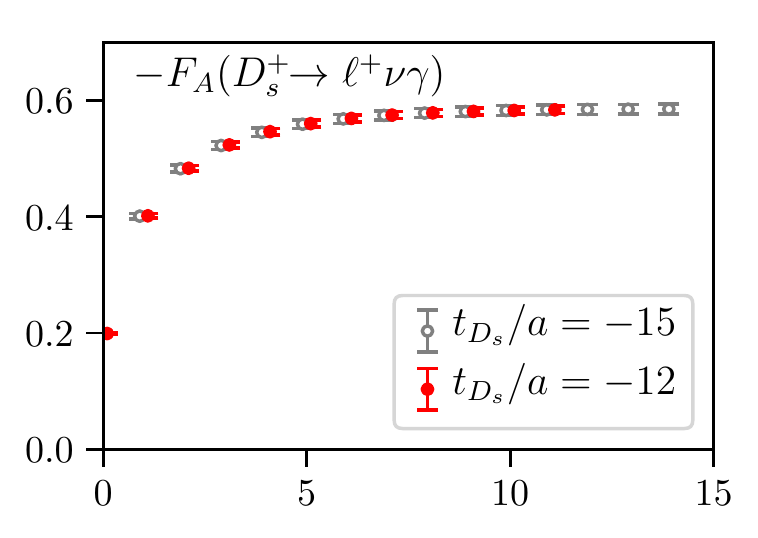}

 \vspace{-4ex}

\hspace{0.5ex}\includegraphics[width=0.47\linewidth]{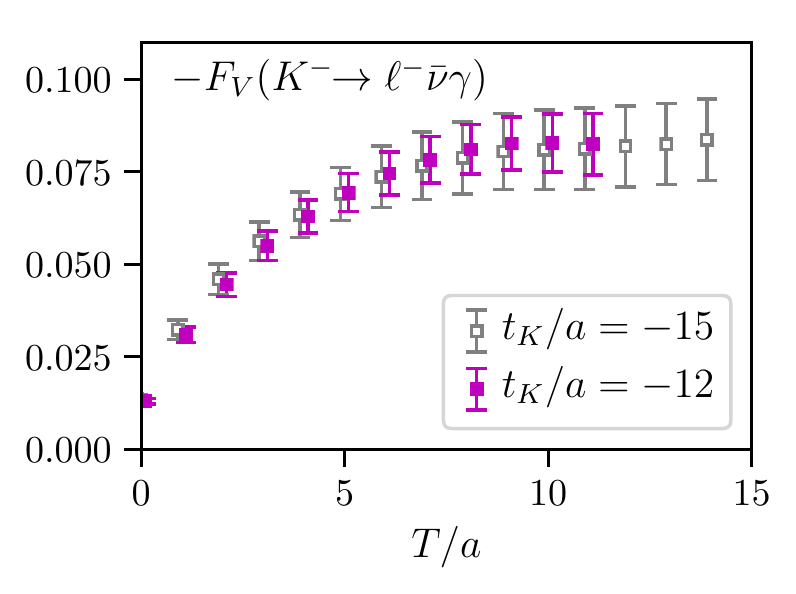} \hfill \includegraphics[width=0.44\linewidth]{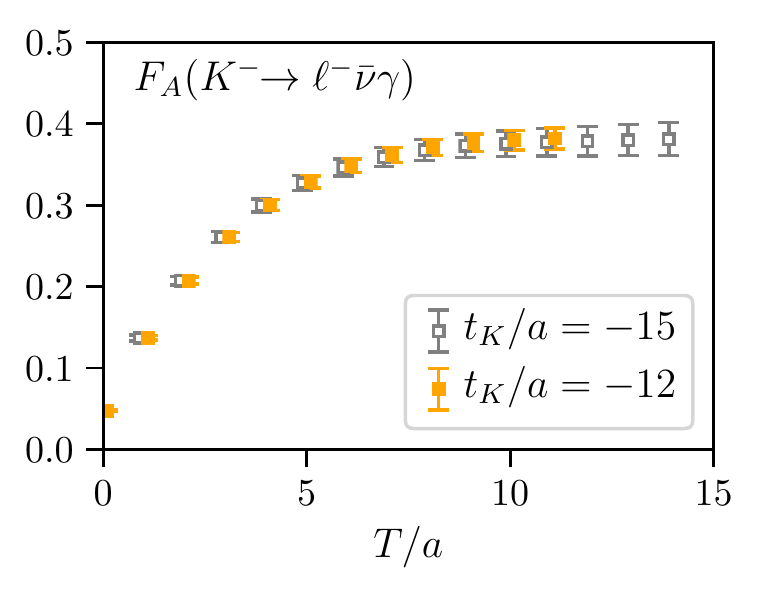}

 \vspace{-2ex}

 \caption{\label{fig:FFvsT}The $D_s^+ \to \ell^+ \nu\gamma$ and $K^- \to \ell^-\bar{\nu}\gamma$ form factors at $\mathbf{p}_\gamma=(0,0,1)\frac{2\pi}{L}$ as a function of the summation range $T$, for two different meson-field insertion times.}
\end{figure}

\begin{figure}[h]

\includegraphics[height=0.23\textheight]{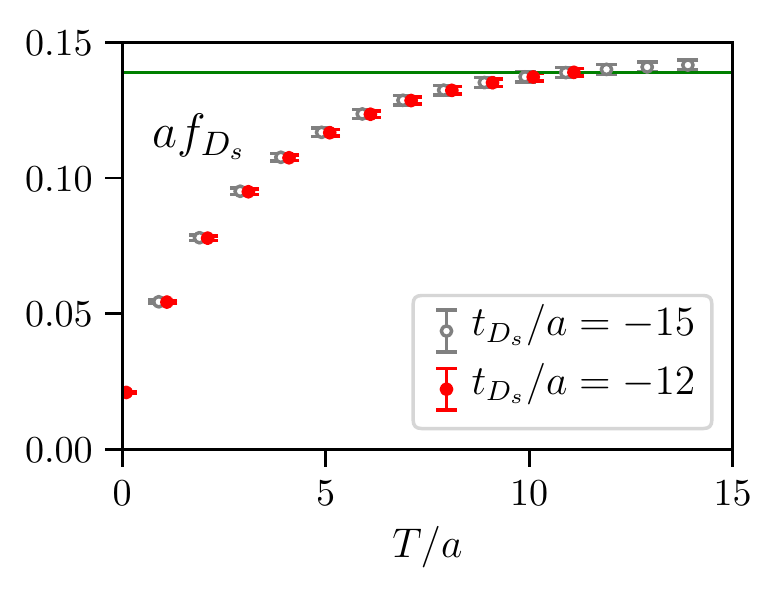} \hfill \includegraphics[height=0.23\textheight]{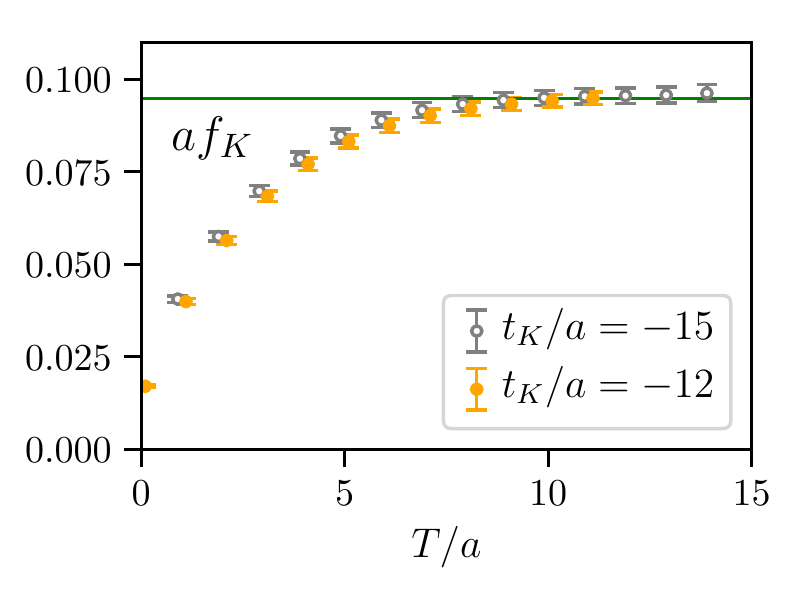}
 
 \vspace{-2.5ex}

 \caption{\label{fig:fvsT}The $D_s$ and $K$ decay constants extracted from $T_{\mu\nu}$ at $\mathbf{p}_\gamma=(0,0,1)\frac{2\pi}{L}$, as a function of the summation range $T$, for two different meson-field insertion times. For the $D_s$, the horizontal line shows the physical value from Ref.~\cite{Aoki:2019cca}. For the $K$, the horizontal line shows the value computed on the same ensemble with the standard method in Ref.~\cite{Aoki:2010dy}.}
\end{figure}

In this section, we present some early numerical results for the $D_s^+ \to \ell^+ \nu\gamma$ and $K^- \to \ell^-\bar{\nu}\gamma$ form factors. These results are from
only 25 configurations of the ``24I'' RBC/UKQCD ensemble \cite{Aoki:2010dy} with $2+1$ flavors of domain-wall fermions and the Iwasaki gauge action, with \mbox{$a^{-1}=1.785(5)$} GeV and $m_\pi=340(1)$ MeV. For the light and strange valence quarks, we use the same domain-wall action as in Ref.~\cite{Aoki:2010dy}. The valence charm quark is implemented with a M\"obius domain-wall action with stout-smeared gauge links ($N=3$, $\rho=0.1$), $L_5/a=12$, $a M_5=1.0$, $a m_f =0.6$ \cite{Boyle:2018knm}, which approximately corresponds to the physical charm-quark mass. We use local currents with ``mostly nonperturbative'' renormalization. Gaussian smearing is performed for the lighter quark in the meson interpolating field.
We start with a $\mathbb{Z}_2$ random-wall source at the time slice of the weak current (denoted as time ``0'' here) and perform sequential inversions through the meson interpolating field; disconnected diagrams are presently neglected. All-mode averaging \cite{Shintani:2014vja} with 16 sloppy and 1 exact samples per configuration is employed; the 16 sloppy samples correspond to 16 different starting time slices. Our initial calculations used $\mathbf{p}_{K/D_s}=0$ and $\mathbf{p}^2_\gamma\in \{1,2,3,4,5\} \left(\frac{2\pi}{L}\right)^2$.

\begin{figure}[t]

 \vspace{-3ex}
 
 \includegraphics[height=0.23\textheight]{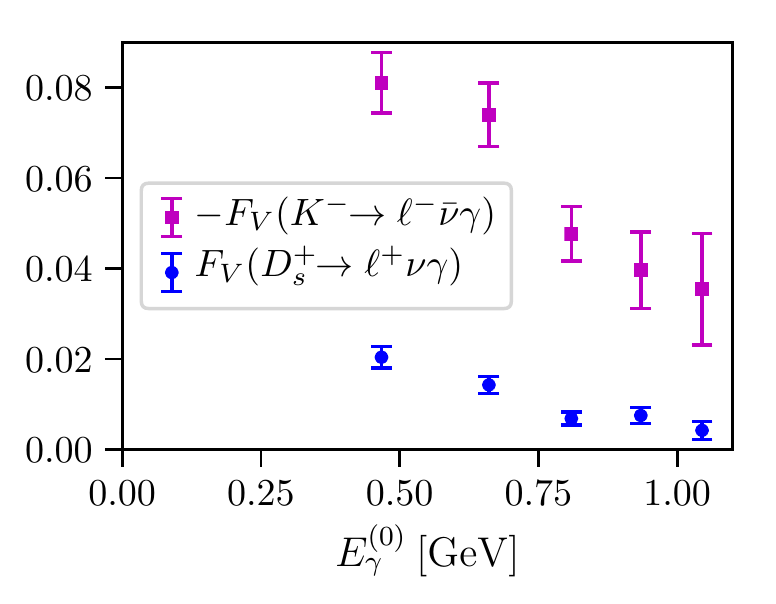} \hfill \includegraphics[height=0.23\textheight]{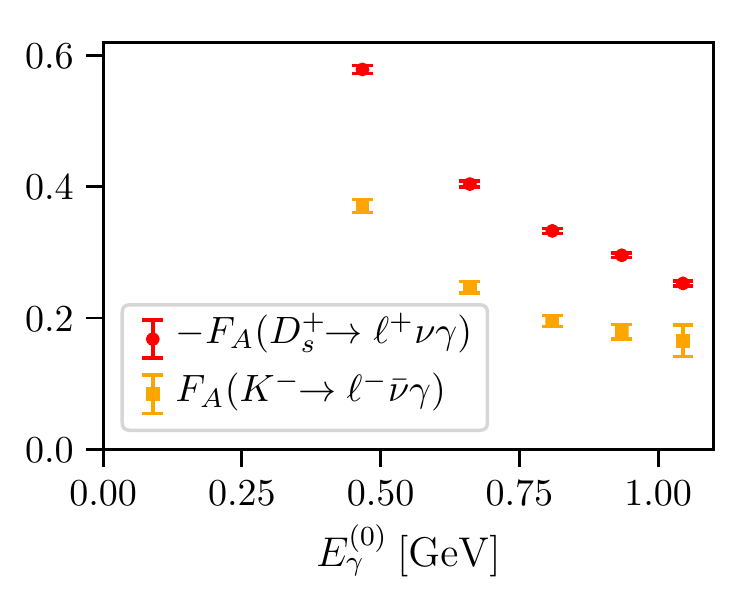}
 
 \vspace{-2.5ex}
 
 \caption{\label{fig:FFvsEgamma}The $D_s^+ \to \ell^+ \nu\gamma$ and $K^- \to \ell^-\bar{\nu}\gamma$ form factors as a function of the photon energy. The results shown here
 were obtained with $T/a=8$ and $t_{K/D_s}/a=-12$. Only the statistical uncertainties are given.}
\end{figure}

Figure \ref{fig:Cmunu} shows examples of the $D_s^+ \to \ell^+ \nu\gamma$ three-point functions. Multiplying by $e^{E_\gamma t}$ and summing over $t$ gives $T_{\mu\nu}$ for sufficiently large summation range $T$. The form factors $F_V$ and $F_A$ extracted from $T_{\mu\nu}$ (at the lowest photon momentum) are shown as a function of $T$ in Fig.~\ref{fig:FFvsT}. The results plateau at approximately $T/a=8$. We also extracted the meson decay constants from the $v^\mu v^\nu$ term in $T_{\mu\nu}$. As can be seen in Fig.~\ref{fig:fvsT}, the results agree with the known values, which is a valuable test of our calculation. Finally, Fig.~\ref{fig:FFvsEgamma} shows the form factors $F_V$ and $F_A$ as a function of the photon energy. Note that, with our current choice of momenta, all of the photon energies are above the physical region for $K^- \to \ell^-\bar{\nu}\gamma$. The results for $F_A$ are dominated by the point-like contribution equal to $- e_\ell f_{K/D_s}/E_\gamma^{(0)}$.

\section{Conclusions and Outlook}

We have shown that the form factors describing radiative leptonic decays can be calculated on the lattice; even though they involve a nonlocal matrix element, the use of imaginary time poses no difficulty in this case. The early results shown here for $D_s^+ \to \ell^+ \nu\gamma$ and $K^- \to \ell^-\bar{\nu}\gamma$ cover photon energies from approximately 0.5 to 1 GeV. For $K^- \to \ell^-\bar{\nu}\gamma$ we need to reach lower photon energies to compare with experiment; this can be achieved by using moving frames (i.e., nonzero $\mathbf{p}_K$) and/or a larger volume. To study the $B_{(s)}$ radiative leptonic decays with the domain-wall action for the heavy quark, we will need to extrapolate in the mass. We are also considering calculations directly at the physical $b$-quark mass using the ``relativistic heavy-quark action'' \cite{Christ:2006us}, but,
because this action is only on-shell improved, additional steps are likely needed to remove unphysical behavior occurring when the electromagnetic and weak currents get close to each other.

\vspace{1ex}

\noindent \textbf{Acknowledgments:} We thank the RBC and UKQCD Collaborations for providing the gauge-field configurations. C.K.\ and S.M.\ are supported by the US DOE, Office of Science, Office of HEP under award number DE{-SC}0009913. S.M.\ is also supported by the RIKEN BNL Research Center. A.S. and C.L. are supported in part by US DOE Contract No. DESC0012704(BNL). During a part of this work, C.L. was also supported by a DOE Office of Science Early Career Award. This work used resources at TACC that are part of XSEDE, supported by NSF grant number ACI-1548562.

\linespread{0.93}

\setlength{\itemsep}{0pt}

\providecommand{\href}[2]{#2}\begingroup\raggedright\endgroup


\begin{thebibliography}{10}

\bibitem{Aebischer:2019mlg}
J.~Aebischer, W.~Altmannshofer, D.~Guadagnoli, M.~Reboud, P.~Stangl, and D.~M.
  Straub, ``{$B$-decay discrepancies after Moriond 2019},''
\href{http://arxiv.org/abs/1903.10434}{{\ttfamily arXiv:1903.10434}}.
%%CITATION = ARXIV:1903.10434;%%.

\bibitem{Atwood:1994za}
D.~Atwood, G.~Eilam, and A.~Soni, ``{Pure leptonic radiative decays $B^\pm, D_s
  \to \ell\nu\gamma$ and the annihilation graph},''
  \href{http://dx.doi.org/10.1142/S0217732396001090}{Mod. Phys. Lett.
  {\bfseries A11} (1996) 1061},
\href{http://arxiv.org/abs/hep-ph/9411367}{{\ttfamily arXiv:hep-ph/9411367}}.
%%CITATION = HEP-PH/9411367;%%.

\bibitem{Korchemsky:1999qb}
G.~P. Korchemsky, D.~Pirjol, and T.-M. Yan, ``{Radiative leptonic decays of B
  mesons in QCD},'' \href{http://dx.doi.org/10.1103/PhysRevD.61.114510}{Phys.
  Rev. {\bfseries D61} (2000) 114510},
\href{http://arxiv.org/abs/hep-ph/9911427}{{\ttfamily arXiv:hep-ph/9911427}}.
%%CITATION = HEP-PH/9911427;%%.

\bibitem{DescotesGenon:2002mw}
S.~Descotes-Genon and C.~T. Sachrajda, ``{Factorization, the light cone
  distribution amplitude of the B meson and the radiative decay $B \to \gamma
  \ell \nu_\ell$},''
  \href{http://dx.doi.org/10.1016/S0550-3213(02)01066-0}{Nucl. Phys. {\bfseries
  B650} (2003) 356},
\href{http://arxiv.org/abs/hep-ph/0209216}{{\ttfamily arXiv:hep-ph/0209216}}.
%%CITATION = HEP-PH/0209216;%%.

\bibitem{Lunghi:2002ju}
E.~Lunghi, D.~Pirjol, and D.~Wyler, ``{Factorization in leptonic radiative $B
  \to \gamma e \nu$ decays},''
  \href{http://dx.doi.org/10.1016/S0550-3213(02)01032-5}{Nucl. Phys. {\bfseries
  B649} (2003) 349},
\href{http://arxiv.org/abs/hep-ph/0210091}{{\ttfamily arXiv:hep-ph/0210091}}.
%%CITATION = HEP-PH/0210091;%%.

\bibitem{Braun:2012kp}
V.~M. Braun and A.~Khodjamirian, ``{Soft contribution to $B\to \gamma \ell
  \nu_\ell$ and the $B$-meson distribution amplitude},''
  \href{http://dx.doi.org/10.1016/j.physletb.2012.11.047}{Phys. Lett.
  {\bfseries B718} (2013) 1014},
\href{http://arxiv.org/abs/1210.4453}{{\ttfamily arXiv:1210.4453}}.
%%CITATION = ARXIV:1210.4453;%%.

\bibitem{Beneke:2018wjp}
M.~Beneke, V.~M. Braun, Y.~Ji, and Y.-B. Wei, ``{Radiative leptonic decay $B\to
  \gamma \ell \nu_\ell$ with subleading power corrections},''
  \href{http://dx.doi.org/10.1007/JHEP07(2018)154}{JHEP {\bfseries 07} (2018)
  154},
\href{http://arxiv.org/abs/1804.04962}{{\ttfamily arXiv:1804.04962}}.
%%CITATION = ARXIV:1804.04962;%%.

\bibitem{Wang:2018wfj}
Y.-M. Wang and Y.-L. Shen, ``{Subleading-power corrections to the radiative
  leptonic $B \to \gamma \ell \nu$ decay in QCD},''
  \href{http://dx.doi.org/10.1007/JHEP05(2018)184}{JHEP {\bfseries 05} (2018)
  184},
\href{http://arxiv.org/abs/1803.06667}{{\ttfamily arXiv:1803.06667}}.
%%CITATION = ARXIV:1803.06667;%%.

\bibitem{Beneke:1999br}
M.~Beneke, G.~Buchalla, M.~Neubert, and C.~T. Sachrajda, ``{QCD factorization
  for $B \to \pi \pi$ decays: Strong phases and CP violation in the heavy quark
  limit},'' \href{http://dx.doi.org/10.1103/PhysRevLett.83.1914}{Phys. Rev.
  Lett. {\bfseries 83} (1999) 1914},
\href{http://arxiv.org/abs/hep-ph/9905312}{{\ttfamily arXiv:hep-ph/9905312}}.
%%CITATION = HEP-PH/9905312;%%.

\bibitem{Gelb:2018end}
{\bfseries Belle} Collaboration, M.~Gelb {\em et~al.}, ``{Search for the rare
  decay of $B^+ \to \ell^{\,+} \nu_{\ell} \gamma$ with improved hadronic
  tagging},'' \href{http://dx.doi.org/10.1103/PhysRevD.98.112016}{Phys. Rev.
  {\bfseries D98} (2018) 112016},
\href{http://arxiv.org/abs/1810.12976}{{\ttfamily arXiv:1810.12976}}.
%%CITATION = ARXIV:1810.12976;%%.

\bibitem{Guadagnoli:2017quo}
D.~Guadagnoli, M.~Reboud, and R.~Zwicky, ``{$B_{s}^{0} \to \ell^{+}\ell^{-}
  \gamma$ as a test of lepton flavor universality},''
  \href{http://dx.doi.org/10.1007/JHEP11(2017)184}{JHEP {\bfseries 11} (2017)
  184},
\href{http://arxiv.org/abs/1708.02649}{{\ttfamily arXiv:1708.02649}}.
%%CITATION = ARXIV:1708.02649;%%.

\bibitem{Dettori:2016zff}
F.~Dettori, D.~Guadagnoli, and M.~Reboud, ``{$B^{0}_{s} \to
  \mu^{+}\mu^{-}\gamma$ from $B^{0}_{s} \to \mu^{+}\mu^{-}$},''
  \href{http://dx.doi.org/10.1016/j.physletb.2017.02.048}{Phys. Lett.
  {\bfseries B768} (2017) 163},
\href{http://arxiv.org/abs/1610.00629}{{\ttfamily arXiv:1610.00629}}.
%%CITATION = ARXIV:1610.00629;%%.

\bibitem{Ablikim:2017twd}
{\bfseries BESIII} Collaboration, M.~Ablikim {\em et~al.}, ``{Search for the
  radiative leptonic decay $D^{+}\to \gamma e^{+} {\nu}_{e}$},''
  \href{http://dx.doi.org/10.1103/PhysRevD.95.071102}{Phys. Rev. {\bfseries
  D95} (2017) 071102},
\href{http://arxiv.org/abs/1702.05837}{{\ttfamily arXiv:1702.05837}}.
%%CITATION = ARXIV:1702.05837;%%.

\bibitem{Ablikim:2019bil}
{\bfseries BESIII} Collaboration, M.~Ablikim {\em et~al.}, ``{Search for the
  decay $D_s^+\rightarrow \gamma e^+\nu_e$},''
  \href{http://dx.doi.org/10.1103/PhysRevD.99.072002}{Phys. Rev. {\bfseries
  D99} (2019) 072002},
\href{http://arxiv.org/abs/1902.03351}{{\ttfamily arXiv:1902.03351}}.
%%CITATION = ARXIV:1902.03351;%%.

\bibitem{Tanabashi:2018oca}
{\bfseries Particle Data Group} Collaboration, M.~Tanabashi {\em et~al.},
  ``{Review of Particle Physics},''
\href{http://dx.doi.org/10.1103/PhysRevD.98.030001}{Phys. Rev. {\bfseries D98}
  (2018) 030001}.
%%CITATION = PHRVA,D98,030001;%%.

\bibitem{Soni2018}
A.~Soni, ``{Flavor anomalies \& the lattice},'' 2018.
\newblock Presentation at Lattice 2018, East Lansing, MI, USA.

\bibitem{Martinelli2019}
{\bfseries RM123} Collaboration, G.~Martinelli, ``{Electromagnetic Corrections
  to Decay Amplitudes: Real Emissions in Leptonic Decays},'' 2019.
\newblock Presentation at Lattice 2019, Wuhan, China.

\bibitem{Ji:2001wha}
X.-d. Ji and C.-w. Jung, ``{Studying hadronic structure of the photon in
  lattice QCD},'' \href{http://dx.doi.org/10.1103/PhysRevLett.86.208}{Phys.
  Rev. Lett. {\bfseries 86} (2001) 208},
\href{http://arxiv.org/abs/hep-lat/0101014}{{\ttfamily arXiv:hep-lat/0101014}}.
%%CITATION = HEP-LAT/0101014;%%.

\bibitem{Dudek:2006ut}
J.~J. Dudek and R.~G. Edwards, ``{Two Photon Decays of Charmonia from Lattice
  QCD},'' \href{http://dx.doi.org/10.1103/PhysRevLett.97.172001}{Phys. Rev.
  Lett. {\bfseries 97} (2006) 172001},
\href{http://arxiv.org/abs/hep-ph/0607140}{{\ttfamily arXiv:hep-ph/0607140}}.
%%CITATION = HEP-PH/0607140;%%.

\bibitem{Feng:2012ck}
X.~Feng, S.~Aoki, H.~Fukaya, S.~Hashimoto, T.~Kaneko, J.-i. Noaki, and
  E.~Shintani, ``{Two-photon decay of the neutral pion in lattice QCD},''
  \href{http://dx.doi.org/10.1103/PhysRevLett.109.182001}{Phys. Rev. Lett.
  {\bfseries 109} (2012) 182001},
\href{http://arxiv.org/abs/1206.1375}{{\ttfamily arXiv:1206.1375}}.
%%CITATION = ARXIV:1206.1375;%%.

\bibitem{Aoki:2019cca}
{\bfseries Flavour Lattice Averaging Group} Collaboration, S.~Aoki {\em
  et~al.}, ``{FLAG Review 2019},''
\href{http://arxiv.org/abs/1902.08191}{{\ttfamily arXiv:1902.08191}}.
%%CITATION = ARXIV:1902.08191;%%.

\bibitem{Aoki:2010dy}
{\bfseries RBC/UKQCD} Collaboration, Y.~Aoki {\em et~al.}, ``{Continuum Limit
  Physics from 2+1 Flavor Domain Wall QCD},''
  \href{http://dx.doi.org/10.1103/PhysRevD.83.074508}{Phys. Rev. {\bfseries
  D83} (2011) 074508},
\href{http://arxiv.org/abs/1011.0892}{{\ttfamily arXiv:1011.0892}}.
%%CITATION = ARXIV:1011.0892;%%.

\bibitem{Boyle:2018knm}
{\bfseries RBC/UKQCD} Collaboration, P.~A. Boyle, L.~Del~Debbio, N.~Garron,
  A.~Jüttner, A.~Soni, J.~T. Tsang, and O.~Witzel, ``{SU(3)-breaking ratios
  for $D_{(s)}$ and $B_{(s)}$ mesons},''
\href{http://arxiv.org/abs/1812.08791}{{\ttfamily arXiv:1812.08791}}.
%%CITATION = ARXIV:1812.08791;%%.

\bibitem{Shintani:2014vja}
E.~Shintani, R.~Arthur, T.~Blum, T.~Izubuchi, C.~Jung, and C.~Lehner,
  ``{Covariant approximation averaging},''
  \href{http://dx.doi.org/10.1103/PhysRevD.91.114511}{Phys. Rev. {\bfseries
  D91} (2015) 114511},
\href{http://arxiv.org/abs/1402.0244}{{\ttfamily arXiv:1402.0244}}.
%%CITATION = ARXIV:1402.0244;%%.

\bibitem{Christ:2006us}
N.~H. Christ, M.~Li, and H.-W. Lin, ``{Relativistic Heavy Quark Effective
  Action},'' \href{http://dx.doi.org/10.1103/PhysRevD.76.074505}{Phys. Rev.
  {\bfseries D76} (2007) 074505},
\href{http://arxiv.org/abs/hep-lat/0608006}{{\ttfamily arXiv:hep-lat/0608006}}.
%%CITATION = HEP-LAT/0608006;%%.

\end{thebibliography}
\end{document}